\documentclass[aps,prd,twocolumn,superscriptaddress,longbibliography]{revtex4-1}
\usepackage{amsmath}
\usepackage{amssymb}
\usepackage{graphicx}
\usepackage{tensor}
\usepackage{float}
\usepackage{caption}
\usepackage{color}
\usepackage{hyperref}
\definecolor{DarkGreen}{rgb}{0.1,0.5,0.1}

\usepackage{accents}
\newcommand*{\dt}[1]{%
  \accentset{\mbox{\large\bfseries .}}{#1}}
\newcommand*{\ddt}[1]{%
  \accentset{\mbox{\large\bfseries .\hspace{-0.25ex}.}}{#1}}

\newcommand{\Christ}[3]{\ensuremath{
\left\{ {}\indices{_#2}\!{}\indices{^#1}{}\!\indices{_#3} \right\}
}}

\begin{document}

\newcommand{\bia}[1]{{\bf \textcolor{red}{#1}}}
\newcommand{\mau}[1]{{\bf \textcolor{blue}{#1}}}
\newcommand{\riba}[1]{{\bf \textcolor{DarkGreen}{#1}}}
\newcommand{\luc}[1]{{\bf \textcolor{cyan}{#1}}}

\title {How does light move in a generic metric-affine background?}
%\\Ray optics and
%generalized Etherington's reciprocity relation}
\author{Lucas T. Santana}
\affiliation{Universidade Federal do Rio de Janeiro,
Instituto de F\'\i sica, \\
CEP 21941-972 Rio de Janeiro, RJ, Brazil}
\author{Maur\'\i cio O. Calv\~ao}
\affiliation{Universidade Federal do Rio de Janeiro,
Instituto de F\'\i sica, \\
CEP 21941-972 Rio de Janeiro, RJ, Brazil}
\author{Ribamar R. R. Reis}
\affiliation{Universidade Federal do Rio de Janeiro,
Instituto de F\'\i sica, \\
CEP 21941-972 Rio de Janeiro, RJ, Brazil}
\affiliation{Observat\'orio do Valongo, Universidade Federal do Rio de Janeiro, \\ Ladeira do Pedro Ant\^onio 43, CEP 20080-090 Rio de Janeiro, RJ, Brazil}
\author{Beatriz B. Siffert}
\affiliation{Universidade Federal do Rio de Janeiro,
Instituto de F\'\i sica, \\
CEP 21941-972 Rio de Janeiro, RJ, Brazil}
\affiliation{Universidade Federal do Rio de Janeiro,
P\'olo de Xer\'em, \\
CEP 25245-390 Duque de Caxias, RJ, Brazil}

\begin{abstract}

Light is the richest information retriever for most physical systems,
particularly so for astronomy and cosmology, in which gravitation is of
paramount importance, and also for solid state defects and metamaterials, in 
which some effects can be
mimicked by non-Euclidean or even non-Riemannian geometries. Thus, it is
expedient to probe light motion in geometrical backgrounds alternative to that 
of general relativity. Here we investigate this issue in generic metric-affine theories and
derive (i) the expression, in the geometrical optics (eikonal) limit, for light
trajectories, showing that they still are null (extremal) geodesics and thus, 
in general, no longer autoparallels, (ii) a generic {formula} to obtain
the relation between source (galaxy) and reception (observer) angular size
(area) distances, generalizing Etherington's original distance reciprocity
relation (DRR), and then applying it to two particular
representative non-Riemannian geometries. First
in metric-compatible, completely antisymmetric torsion geometries, the
generalized DRR is not changed at all, 
and then in Weyl integrable spacetimes, the generalized DRR assumes
a specially simple expression. 

\end{abstract}

\maketitle

\section{Introduction} In 1933, at the culmination of a debate on
relativistic distances, Etherington derived relations between two kinds of
distance in an arbitrary
Lorentzian geometry, the so-called 
distance reciprocity and duality relations
\cite{*[][{;
republished in }] Etherington33, *Etherington07, Ellis07}.
This beautiful result, based on properties of null geodesics, lies at the
heart of essentially all observations in astronomy and cosmology, and its
refutation would be ``a catastrophe from the theoretician's viewpoint''
\cite{Kristian66} or ``a major crisis for observational cosmology'' 
\cite{Ellis07}.

The usual distance reciprocity relation (DRR) connects the angular size 
distance, $D_S$, from an arbitrary instantaneous observer at the source to the 
angular size distance, $D_R$, from an arbitrary instantaneous observer at the 
reception (cf. Fig.~\ref{fig:reciprocity}). Its derivation is carried out by 
assuming, besides the Riemannian (in fact, Lorentzian) character of the 
spacetime, that there are neither interruption (absorption or creation) of light 
rays nor bifurcations (birefringence), and it reads $D_S=(1+z)\,D_R$, where $z$ 
is the redshift between the two instantaneous observers.  If, furthermore, the 
energy-momentum tensor of the electromagnetic field is (covariantly) conserved 
(``photons are conserved''), then the so-called luminosity distance, $D_L$, may 
be related to $D_S$ so that we get Etherington's famous usual distance duality 
relation (DDR): $D_L = (1+z)^2\,D_R$\,. We remark that, in a cosmological (or 
even astronomical) setting, in general, none of the three distances are directly 
measurable; we always have to assume or derive the value of some proper feature 
of the inaccessible source (emission beam solid angle, transverse area or 
luminosity). To investigate a possible violation of this canonical DDR, it is 
expedient to define
\begin{equation}
 \eta := \dfrac{D_L}{(1+z)^2D_R}\,. \label{eta}
\end{equation}
{For general relativity (GR), of course $\eta = 1$. Observational constraints
on 
its value} have been extensively explored in the recent literature
\cite{Bassett04,Uzan04,More09,Avgoustidis10,Khedekar11,
Lima11,Nair11,Cardone12,Holanda12,Nair12,Ellis13,Yang13,
SantosdaCosta15,Liao15,Wu15, Avgoustidis16, Rasanen16}.

Light is an electromagnetic phenomenon and thus its trajectories, in the 
geometrical optics or eikonal (high frequency, nearly monochromatic plane wave) 
approximation, should be suitably derived from Maxwell's equations in a
convenient background. Both in the special relativistic context and in GR as 
well, this leads to the well-known and pleasing result that light moves on null 
(extremal, metric) geodesics (or autoparallels or affine geodesics, which do coincide
with the metric geodesics in a Riemannian geometry) \cite{Born99,Schneider92}.
However, there are many alternative
theories of
gravity or even effective field theories (for metamaterials or solid state 
physics), which are built on top of more general non-Riemannian geometries. We 
will be particularly interested in those where, in contrast to Einstein's GR, 
the affine connection has nonvanishing torsion and nonmetricity (to be defined 
in the next section); for general reviews and motivation, see 
\cite{Hehl95,Hammond02,Shapiro02,Ni10,Vitagliano14}. This is a sufficiently wide 
class of theories to include: Einstein-Cartan theory \cite{Hehl76}, teleparallel 
theories \cite{Aldrovandi13}, Weyl theories \cite{Scholz15}, metric-affine 
gauge theories \cite{Hehl95}, Kalb-Ramond string fields 
\cite{Mukhopadhyaya02,Das14}, metamaterials \cite{Horsley11} and to also 
incorporate a generalized Ehlers-Pirani-Schild approach for chronogeometry 
\cite{Ehlers72}. 

Our aim is twofold: to derive, under the scope of a completely general
metric-affine geometry, (i) the trajectories light follows and, therefrom, (ii) 
the generalized DRR. As an application, we employ it to two particular
non-Riemannian geometries.

\section{General metric-affine background}

The class of theories we envisage are those which have a metric-affine
background, constituted by any model $({\mathcal M}, g_{\alpha\beta},
{\Gamma^\alpha}_{\mu\nu})$, where ${\mathcal M}$ is the base
manifold, $g_{\alpha\beta}$ is a Lorentzian metric tensor (with signature $+2$) and ${\Gamma^\alpha}_{\mu\nu}$ is
a generic (asymmetric) affine connection, such that, in general, the
corresponding torsion and nonmetricity tensors are defined respectively by
\begin{align}
 {T^{\alpha}}_{\mu\nu} &:= 2{\Gamma^\alpha}_{[\mu\nu]}\,, \label{torsion} \\
 Q_{\alpha\beta\gamma} &:= \nabla_{\gamma} g_{\alpha \beta}\,, \label{nonmetricity}
\end{align}
{where, of course, $\nabla$ stands for the covariant derivative with respect to
the fundamental connection $\Gamma$, whereas, later on,
$\widehat{\nabla}$ will stand for the covariant derivative with respect to the (auxiliary) Levi-Civita connection $\widehat{\Gamma}$.
These connections are related by a useful identity \cite{Schouten54}:
}
\begin{equation}
 {\Gamma^\alpha}_{\mu\nu} = {{\widehat\Gamma}^{\alpha}}_{\ \,\mu\nu} +
{\Delta^\alpha}_{\mu\nu}\,,
\end{equation}
where
\begin{align}
{\widehat{\Gamma}}\indices{^\alpha_{\mu\nu}}
&:= \Christ{\alpha}{\mu}{\nu}
:=\dfrac{1}{2}g^{\alpha\beta}{\left( \partial_{\nu}g_{\beta\mu} + \partial_{\mu}g_{\beta\nu} -
\partial_{\beta}g_{\mu\nu} \right)}\,,  \\
 {\Delta^\alpha}_{\mu\nu} &:= {K^\alpha}_{\mu\nu} + {D^\alpha}_{\mu\nu}\,,
\end{align}
and
\begin{align}
 {K^\alpha}_{\mu\nu} &:=  \dfrac{1}{2}\left( {T_{\mu\nu}}^\alpha +
{T_{\nu\mu}}^\alpha + {T^\alpha}_{\mu\nu} \right)\,,  \\
 {D^\alpha}_{\mu\nu} &:= \dfrac{1}{2} \left( Q\indices{_{\mu\nu}^\alpha} -
Q\indices{^\alpha_{\mu\nu}} - Q\indices{_\nu^\alpha_\mu} \right)\,.
\end{align}
Here $\Christ{\alpha}{\mu}{\nu}$ is the Christoffel
symbol (of the second kind),
$K\indices{^\alpha_{\mu\nu}}$ is the contortion tensor and
$D\indices{^\alpha_{\mu\nu}}$ is the deflection tensor. Of course, the usual
Lorentzian case, which
includes GR and all $f(R)$ theories, corresponds to 
${T^\alpha}_{\mu\nu} = Q_{\alpha\beta\mu} = 0$\,, {whence
${\Gamma^\alpha}_{\mu\nu}={\widehat{\Gamma}}\indices{^\alpha_{\mu\nu}}$\,.} Our results are
independent of any specific form for the governing equations of {the
fundamental gravitational fields, which, without any loss of generality, will be taken as 
${\boldsymbol {\mathcal G}}:=(g\indices{_{\alpha\beta}},
T\indices{^\alpha_{\mu\nu}},
Q\indices{_{\alpha\beta\gamma}})$\,. }

\section{Geometrical optics approximation}

There are several classical approaches aiming to derive the trajectories
followed by light, in the geometrical optics or eikonal approximation:
asymptotic series \cite{Ehlers67, Misner73, Schneider92, Perlick00, Bona11}, 
Fourier transform \cite{Born99, Horsley11}
and characteristics or discontinuities \cite{Courant89, Kline65, Born99,
Friedlander75, Bona11}.

Here we follow the asymptotic series one and, therefore, we only have to impose
conditions on the higher-order derivative terms (the principal part) for the
generalized vacuum Maxwell equations, in the absence of sources, of the
antisymmetric electromagnetic
field tensor, $F\indices{_{\alpha\beta}}$. Inspired by the usual case, we assume
they are still two sets of first-order (in $F\indices{_{\alpha\beta}}$)
linear homogeneous partial
differential equations given by 
\begin{align}
\left[\nabla_{\beta} + \chi_{\beta}({\boldsymbol{\mathcal G}},\nabla{\boldsymbol{\mathcal G}},\ldots)\right]F^{\alpha\beta} &=0\,,\label{Maxwell_1} \\
\left[\nabla_{[\alpha}+\zeta_{[\alpha}({\boldsymbol{\mathcal G}},\nabla{\boldsymbol{\mathcal G}},\ldots)\right]F_{\beta\gamma]} &=0\,.\label{Maxwell_2}
\end{align}
Here $\chi_{\alpha}$ and $\zeta_{\alpha}$ are arbitrary covariant vector 
fields dependent only on ${\boldsymbol {\mathcal G}}$ and their (covariant) 
derivatives up to a finite order.
Constraints on their expressions might be established either from additional physical
assumptions, such as the existence of a 4-potential or charge conservation, or from a
variational approach.
Of course, the existence of a 4-potential will impose, through Poincar\'e's lemma, a constraint on $\zeta_{\alpha}$
whereas charge conservation will restrict $\chi_{\alpha}$, from Eq. (\ref{Maxwell_1}) with a source term. Hence, if one wants to ensure both, one does not necessarily need to postulate the usual set of Maxwell equations of GR, neither a Riemannian background.

Resuming now our main derivation, we look for (asymptotic) solutions of the generalized Maxwell equations
(\ref{Maxwell_1}) and (\ref{Maxwell_2}) in the form of a monochromatic 
wave:
\begin{equation}\label{plane}
	F_{\mu \nu} = \Re\,{[A_{\mu \nu}(x)e^{iS(x)/\epsilon}]},
\end{equation}
where $ A_{\mu \nu} $ is an antisymmetric tensor field, $ S $ is a real scalar field, the
\textit{phase} of the wave, $ \epsilon $ is a control parameter for
the wavelength, and $\Re$ indicates that the real part of the following expression is to be taken.

Inserting Eq. (\ref{plane}) into Eqs. (\ref{Maxwell_1}) and (\ref{Maxwell_2}), and
imposing the condition $\epsilon\rightarrow 0$, we obtain
\begin{equation}\label{null}
	k^{\mu}k_{\mu} = 0, \hskip 1cm k_{\mu} := \partial_{\mu} S,
\end{equation}
where $k_{\mu}$ is the wave 4-vector, whose integral curves are to be considered
as the light rays. This condition is completely independent of $\chi_{\alpha}$ 
and $\zeta_{\alpha}$. From Eq. (\ref{null}), {we immediately derive our first
simple general 
result, valid for any metric-affine theory and linear generalized Maxwell's 
equations:} the light rays are (extremal) metric geodesics 
($k^\nu\widehat{\nabla}_\nu k^\mu=0\,$), as in GR, although, in general, no 
longer affine geodesics (autoparallels), 
\begin{equation}
\label{light-curves}
	\dt{k}^{\mu} = [T_{\alpha \beta}^{\hskip 0.3cm \mu} + 
	(1/2)Q_{\alpha \beta}^{\hskip 0.4cm \mu} - Q^{\mu}_{\hskip 0.2cm \alpha 
\beta}]k^{\alpha}k^{\beta}\,,
\end{equation}
{in contrast to GR. Here and later on $\dt{Z}^{\alpha \cdots}_{\hskip 0.4cm \beta \cdots}\!\! :=
k^{\nu}\nabla_{\nu}Z^{\alpha \cdots}_{\hskip 0.4cm \beta \cdots}$\,.}

In addition, completing the geometrical optics 
limit, we were able to obtain evolution equations for the scalar amplitude 
as well as the polarization of the electromagnetic wave, showing that both 
quantities are parallely propagated along the light rays, although the family of 
such quantities satisfying such conditions is dependent on the choice of 
$\chi_{\alpha}$ and $\zeta_{\alpha}$.

\section{Generalized distance reciprocity relation} 

Now, from Eq. (\ref{light-curves}), {we derive our second general result:} 
for any metric-affine geometry, the generalized deviation equation for the light 
rays is
\begin{equation}\label{deviation}
	\ddt{X}^{\alpha} = 
	{R^{\alpha}}_{\mu \nu \sigma} k^{\mu} k^{\nu} X^{\sigma} + 
X^{\nu} \nabla_{\nu} \dt{k}^{\alpha} - ({T^{\alpha}}_{\mu
\nu} 
k^{\mu} X^{\nu})\!\dt{\phantom{x}},
\end{equation}
where $X^{\alpha}$ is any connecting vector field associated to a congruence of 
light rays. We stress that 
this result turns out to coincide with the one in \cite{Swaminarayan83}, 
but there it was proven only for the particular case of vanishing nonmetricity.

Next, let $\mathcal{B}_i$, $i=\mathcal{S},\mathcal{R}$, be two 2-parameter infinitesimal 
pencil beams of generalized light rays, such that their vertices are the events 
$i=\mathcal{S}$ (for source) and $i= \mathcal{R}$ (for reception), along a single 
common curve, the so-called fiducial light ray $\mathcal{C}$.
\begin{figure}[H]
   \centering
   \includegraphics[scale = 0.3]{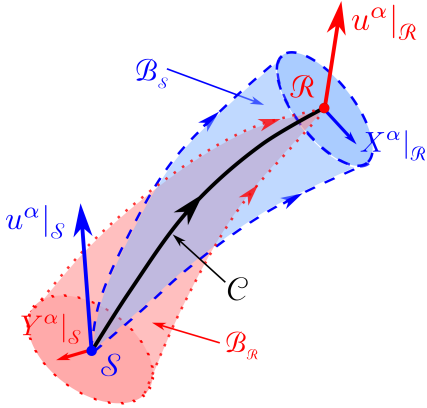}
   \caption{Spacetime diagram for the infinitesimal light beams $\mathcal{B}_{{}_\mathcal{S}}$ (dashed) and
$\mathcal{B}_{{}_\mathcal{R}}$ (dotted) based, respectively, at events source
$\mathcal{S}$ and reception $\mathcal{R}$, their common fiducial light ray
$\mathcal{C}$ (solid), and the corresponding connecting vectors
$\left.X^\alpha\right|_{\mathcal{R}}$ and
$\left.Y^\alpha\right|_{\mathcal{S}}$. }
   \label{fig:reciprocity}
\end{figure}
In the Riemmanian case, if $X^{\mu}$ and $Y^{\mu}$ are the connecting vector fields of $\mathcal{B}_R$ and $\mathcal{B}_S$, respectively, there is a conserved quantity along the fiducial light ray, namely,
\begin{equation}\label{pre-constant}
	Y_{\mu}\dt{X}^{\mu} - X_{\mu}\dt{Y}^{\mu} = \text{constant along } \mathcal{C}.
\end{equation}

Given any pair of connecting vectors of $\mathcal{B}_{{}_\mathcal{S}}$ ($X^{\mu}$ and $\tilde{X}^{\mu}$), and any pair of connecting vectors of $\mathcal{B}_{{}_\mathcal{R}}$ ($Y^{\mu}$ and $\tilde{Y}^{\mu}$), from Eq. (\ref{pre-constant})
\begin{equation}\label{constant-of-motion-riemannian}
	\left.\left(Y_{\mu}\dt{X}^{\mu}\right)\right|_{\mathcal{S}}\left.\left(\tilde{Y}_{\nu}\dt{\tilde{X}}^{\nu}\right)\right|_{\mathcal{S}} = \left.\left(X_{\mu}\dt{Y}^{\mu}\right)\right|_{\mathcal{R}}\left.\left(\tilde{X}_{\nu}\dt{\tilde{Y}}^{\nu}\right)\right|_{\mathcal{R}}.
\end{equation}

Provided that $X^{\mu}$ and $\tilde{X}^{\mu}$ are a pair of orthogonal connecting vector fields
of $\mathcal{B}_{{}_\mathcal{S}}$, belonging to the screen space of $\left.u^\alpha\right|_{\mathcal{R}}$, $Y^{\mu}$  and $\tilde{Y}^{\mu}$, a pair of orthogonal connecting vector fields of $\mathcal{B}_{{}_\mathcal{R}}$, belonging to the screen space of $\left.u^\alpha\right|_{\mathcal{S}}$, the usual DRR, which holds for arbitrary Lorentzian {spacetimes}, is essentially equivalent to 
\cite{Ellis71,Plebanski06} (notice however the different notation)
\begin{equation}\label{drr-relations-riemann}
	dA_{\mathcal{S}}\,d\Omega_{\mathcal{S}}\!\left.(k_\mu
u^\mu)^2\right|_{\mathcal{S}} = 
	dA_{\mathcal{R}}\,d\Omega_{\mathcal{R}}\!\left.(k_\mu
u^\mu)^2\right|_{\mathcal{R}}\,,
\end{equation}
where $dA_{\mathcal{S},\mathcal{R}}$ is an infinitesimal area of the beam 
$\mathcal{B}_{\mathcal{R},\mathcal{S}}$
and
$d\Omega_{\mathcal{R},\mathcal{S}}$ is the corresponding infinitesimal solid 
angle, 
both with respect to the instantaneous observer at the  event given by the
subindex
(cf.\
Fig.~\ref{fig:reciprocity})\,. 

{When reading the classical works on this subject \cite{Ellis71,Plebanski06}, one might be tempted to think that Eq. (\ref{pre-constant}) is a necessary result for the imposition of the previous conditions on the connecting vectors. We, however, follow a different approach, treating the previous constraints on those vectors simply as the initial conditions for their respective set of deviation equations [cf. (\ref{deviation})]. Thus, we see Eq. (\ref{pre-constant}) as a means to relate cosmological observables in $\mathcal{S}$ to their respective counterparts in $\mathcal{R}$.

{Now}, from the usual definitions of redshift, 
$1 + z := \left.(k_\mu u^\mu)\right|_{\mathcal{S}}/\left.(k_\mu u^\mu)\right|_{\mathcal{R}}$\,, 
and the angular size distances, 
$D_{S,R}:=\sqrt{dA_{\mathcal{R},\mathcal{S}}/d\Omega_{\mathcal{S},\mathcal{R}}}
$\,, we immediately have
\begin{equation}\label{drr-riemann}
D_S = (1+z)D_R\,.
\end{equation}

{In a generic metric-affine theory, Eq. (\ref{constant-of-motion-riemannian}) is
{replaced instead by} 
\begin{equation}\label{drr-relations-metric-affine}
\begin{split}
	& \left.\left(Y_{\mu}\dt{X}^{\mu}\right)\right|_{\mathcal{S}}\left.\left(\tilde{Y}_{\nu}\dt{\tilde{X}}^{\nu}\right)\right|_{\mathcal{S}} = \left.\left(X_{\mu}\dt{Y}^{\mu}\right)\right|_{\mathcal{R}}\left.\left(\tilde{X}_{\nu}\dt{\tilde{Y}}^{\nu}\right)\right|_{\mathcal{R}} + \\ 
	& \frac{1}{2}\left[\big(\left.Y_{\rho}\dt{X}^{\rho}\right|_{\mathcal{S}} -\left.X_{\rho}\dt{Y}^{\rho}\right|_{\mathcal{R}}\big) \tilde{\mathcal{I}} +
\big(\left.\tilde{Y}_{\rho}\dt{\tilde{X}}^{\rho}\right|_{\mathcal{S}}-\left.\tilde{X}_{\rho}\dt{\tilde{Y}}^{\rho}\right|_{\mathcal{R}}\big)
\mathcal{I}\right]\,.
\end{split}
\end{equation}
Here $\mathcal{I}$ and $\tilde{\mathcal{I}}$ stand for the functionals 
\begin{align}
	\mathcal{I} & := \int_{\mathcal{C}_{\mathcal{S} \to \mathcal{R}}}
M({\boldsymbol {\mathcal{G}}}, k^\alpha, X^\alpha,
Y^\alpha)\, d\vartheta, \\
	\tilde{\mathcal{I}} & := \int_{\mathcal{C}_{\mathcal{S} \to
\mathcal{R}}} M({\boldsymbol {\mathcal G}},
k^\alpha, \tilde{X}^\alpha, \tilde{Y}^\alpha)\, d\vartheta,
\end{align}
where $\vartheta$ is a parameter along the fiducial light ray, and
\begin{widetext}
\begin{equation}
\label{M-integrand}
\begin{split}
	M := &\left[Q_{\sigma\mu[\nu;\rho]} + Q_{\mu\rho[\sigma;\nu]} + Q_{\mu \nu [\sigma;\rho]} + Q_{\sigma\rho[\mu;\nu]} + Q_{\sigma\nu[\rho;\mu]} + T^{\lambda}_{\hskip 0.2cm \nu [\rho} Q_{\mu]\sigma \lambda} + T^{\lambda}_{\hskip 0.2cm \sigma [\nu} Q_{\rho]\mu \lambda} + T^{\lambda}_{\hskip 0.2cm \mu [\sigma} Q_{\rho]\nu \lambda} + T^{\lambda}_{\hskip 0.2cm \nu \rho;(\sigma}g_{\mu)\lambda} + \right.\\ & \left. T^{\lambda}_{\hskip 0.2cm \nu \rho}T_{(\sigma \mu) \lambda} + 2g_{\lambda(\sigma}T^{\lambda}_{\hskip 0.2cm \mu)[\nu;\rho]} + 2T^{\lambda}_{\hskip 0.2cm [\nu|(\sigma}T_{\mu)\lambda|\rho]} \right]Y^{\rho}k^{\sigma}k^{\mu}X^{\nu} +
X_{\rho}({T^{\rho}}_{\mu\nu} k^{\mu} Y^{\nu})\!\dt{\phantom{x}} -
Y_{\rho}({T^{\rho}}_{\mu\nu} k^{\mu} X^{\nu})\!\dt{\phantom{x}} + \dt{Y}_{\rho}\dt{X}^{\rho} - \\ & \dt{X}_{\rho}\dt{Y}^{\rho} + (X^{\mu}Y_{\nu} - Y^{\mu}X_{\nu})(\dt{k}^{\nu})_{; \mu}.
\end{split}
\end{equation}
\end{widetext}
Here for brevity $ Z^{\alpha \cdots}_{\hskip 0.4cm \beta \cdots ; \mu} := \nabla_{\mu} Z^{\alpha \cdots}_{\hskip 0.4cm \beta \cdots} $.

We can rearrange Eq. (\ref{drr-relations-metric-affine}) (cf. the comments before Eq. (\ref{drr-relations-riemann}) and Eq. (\ref{drr-riemann})) in order to obtain a generalization of the usual DRR, viz.:}
\begin{equation}\label{drr-metric-affine}
	D_S = (1 + z)D_R\big(1 +
\mathcal{J}\big)^{1/2}\,,
\end{equation}
where
\begin{equation}\label{J}
\mathcal{J} := -\frac{\big(\left.Y_{\rho}\dt{X}^{\rho}\right|_{\mathcal{S}} -\left.X_{\rho}\dt{Y}^{\rho}\right|_{\mathcal{R}}\big) \tilde{\mathcal{I}} +
\big(\left.\tilde{Y}_{\rho}\dt{\tilde{X}}^{\rho}\right|_{\mathcal{S}}-\left.\tilde{X}_{\rho}\dt{\tilde{Y}}^{\rho}\right|_{\mathcal{R}}\big)
\mathcal{I}}{2\left.\left(Y_{\mu}\dt{X}^{\mu}\right)\right|_{\mathcal{S}}\left.\left(\tilde{Y}_{\nu}\dt{\tilde{X}}^{\nu}\right)\right|_{\mathcal{S}}}.
\end{equation}

This gives a definite procedure to obtain corrections of the usual distance reciprocity relation due to modified electrodynamics or gravitation, and provides a theoretical grounding for the phenomenological parameterizations in the literature.

Equations (\ref{drr-relations-metric-affine}) to {(\ref{J})} allow us to calculate
$ \mathcal{J} $ regardless {of} the gravitational
field equations or the full form of the sourceless electromagnetic ones, as long as they can be written
as Eqs. (\ref{Maxwell_1}) and (\ref{Maxwell_2}). Of course $ \mathcal{J} $
vanishes for GR (in fact for any Riemannian geometry). {Despite the general form which the functional $ \mathcal{J} $ may assume, we apply, in the next section, the result in Eq. (\ref{drr-metric-affine}}) to two simple non-Riemannian geometries and discuss their most prominent consequences.

\section{Application: two simple cases}

Now we apply the generalized DRR formula in Eq. {
(\ref{drr-metric-affine})} to a couple of simple non-Riemannian
geometries. First, let us consider a metric connection ($Q_{\alpha\beta\mu} =
0$) whose torsion is completely antisymmetric ($T_{\alpha\beta\mu} =
-T_{\beta\alpha\mu}\Rightarrow T_{[\alpha\beta\mu]} = T_{\alpha\beta\mu}$)\,.
This does not necessarily imply the connection is the Levi-Civita one. However
it does imply, through Eq.
(\ref{light-curves}), that light rays follow both affine geodesics
(autoparallels) as well as (extremal) metric geodesics. This
is just the content of the weak equivalence principle, at least for
nonmassive particles. {Moreover, it is straightforward to
show that $ \mathcal{J} $
vanishes}, and the usual DRR of
Riemannian geometry is preserved in the form of Eq.
(\ref{drr-riemann}). In other words, we have shown that Riemannian
geometry is a sufficient condition for the validity of the usual DRR, but not a
necessary one.

The second case we consider is the Weyl integrable {spacetime} (WIST) nonmetric 
($Q_{\mu \nu \lambda} \ne 0 $) symmetric ({$T^{\lambda}_{\hskip 0.2cm \mu \nu} = 0$}) connection: 
$ Q_{\mu \nu \lambda} = g_{\mu \nu} \partial_{\lambda}\phi $, where $\phi$ is a 
completely arbitrary real scalar field. These conditions imply again that light 
rays follow both affine and metric geodesics, although their affine parameters 
now differ. Furthermore, we have shown that {$ \mathcal{J}
= e^{2(\phi|_{\mathcal{R}} - \phi|_{\mathcal{S}})} - 1$
, and} the DRR can now be cast in the 
following form:
\begin{align}\label{wist-drr}
	D_S = (1 + z) D_R e^{(\phi|_{\mathcal{R}} - \phi|_{\mathcal{S}})}\,.
\end{align}
This shows that a convenient choice of the WIST scalar field \cite{Santana16} will provide 
a derivation of the phenomenological modifications of the usual DRR (or DDR; cf. 
below), ordinarily assumed in a vast class of recent works 
\cite{Bassett04, Lima11, Nair11, Holanda12, Wu15}.

\section{Conclusion}

In this work we focused on the influence of the geometrical structure
of {spacetime} on electromagnetic phenomena, namely, light trajectories in the
geometrical optics approximation and the distance reciprocity relation, 
both in generalized metric-affine 
theories.

We have shown that light rays still follow metric geodesics, as in 
GR, although those curves are no longer autoparallels when considering nonvanishing
torsion and nonmetricity. This result holds for any set of partial linear homogeneous
differential equations for the electromagnetic field [cf. Eqs. (\ref{Maxwell_1}) and 
(\ref{Maxwell_2})]. Naturally, if one wishes to ensure the usual symmetries of electromagnetism,
such as charge conservation, or existence of four-potential (gauge symmetry), 
this would imply constraints on $\chi_{\alpha}$ and $\zeta_{\alpha}$ {[cf. the comments after Eqs. (\ref{Maxwell_1}) and (\ref{Maxwell_2})].}

We also obtained that the deviation equation (\ref{deviation})
holds for a more general context, one with arbitrary nonmetricity. Using the previous result,
we were able to exhibit the modifications of DRR in the presence of torsion and nonmetricity, {providing two simple cases as an application}.
We emphasize that these results are completely independent of the field equations
for gravitation.

To obtain a generalized DDR from our DRR, 
Eq.~(\ref{drr-relations-metric-affine}), based solely on the arbitrary 
metric-affine background and our reasonably general, but unspecified, Maxwell's 
equations (\ref{Maxwell_1}) and (\ref{Maxwell_2}), does not seem feasible, 
unless we postulate a conservation of ``photon number.'' 
If we do so, then it is straightforward to show that the parameter $\eta$
from Eq. (\ref{eta}) is related to the functional
$\mathcal{J}$ from Eq.
(\ref{drr-relations-metric-affine}) as $\eta^2 = 1 +
{\mathcal{J}}$\,.
In 
general, however, it is physically transparent that, due to the arbitrariness of 
the interaction between the gravitational fields and the electromagnetic one, the 
most we can hope for is a balance equation for photon number 
\cite{Calvao92}; alternatively, we do not have an explicit expression for the 
electromagnetic energy-momentum tensor field. To establish such a generalized 
balance equation for the photon number one might proceed in three distinct 
ways: to choose an electromagnetic Lagrangian or explicit expressions for 
$\chi_\alpha$ and $\zeta_\alpha$, to follow a thermodynamic approach 
\cite{Calvao92}, or to deal with a kinetic treatment \cite{Lima14}. We tackle 
this issue in a future investigation where observational constraints on specific 
models are explored as well.

\section*{Acknowledgments}
B.B.S. would like to thank Brazilian funding agency CAPES for post doc 
fellowship PNPD Institucional 2940/2011. L.T.S. would like to thank 
Funda\c{c}\~ao Carlos Chagas Filho de Amparo \`a Pesquisa do Estado do Rio de 
Janeiro (FAPERJ) for undergraduate {fellowship} 207137/2015.

\bibliography{light_motion}

\end{document}